\documentclass[report,11pt]{article}

\usepackage{scrextend}
\usepackage{amsmath,amssymb,epsfig,bbm}
\usepackage{MnSymbol}
\usepackage{slashed}
\usepackage{amsthm}

\usepackage{float}

\usepackage{enumerate}

\usepackage{color}
\usepackage{mathrsfs}

\usepackage{dsfont}

\definecolor{co}{cmyk}{0,0.7,0.3,0}
\definecolor{darkgreen}{cmyk}{1,0,1,.2}
\definecolor{m}{rgb}{1,0.1,1}

\newcommand{\be}{\begin{equation}}
\newcommand{\ba}{\begin{eqnarray}}
\newcommand{\ea}{\end{eqnarray}}
\newcommand{\nn}{\nonumber}

\def\d{\delta}

\def\k{\kappa}

\def\p{\pi}

\def\ca{{\cal A}}

\def\ch{{\cal H}}

\def\cm{{\cal M}}

\makeatletter
\newcommand{\eqnum}{\refstepcounter{equation}\textup{\tagform@{\theequation}}}
\makeatother

\newcommand{\pa}{\partial}

\newtheorem*{definition*}{Definition}

\fontfamily{yfrak}

\begin{document}

\vskip 25mm

\begin{center}

{\large\bfseries

Dirac Operators on Configuration Spaces and\\ Yang-Mills Quantum Field Theory

}

\vskip 6ex

Johannes \textsc{Aastrup}$^{a}$\footnote{email: \texttt{aastrup@math.uni-hannover.de}} \&
Jesper M\o ller \textsc{Grimstrup}$^{b}$\footnote{email: \texttt{jesper.grimstrup@gmail.com}}\\ 
\vskip 3ex

$^{a}\,$\textit{Mathematisches Institut, Universit\"at Hannover, \\ Welfengarten 1, 
D-30167 Hannover, Germany.}
\\[3ex]
$^{b}\,$\textit{Copenhagen, Denmark.}
\\[3ex]

{\footnotesize\it This work is financially supported by entrepreneur Kasper Gevaldig, Denmark,\\ and by Master of Science in Engineering Vladimir Zakharov, Granada, Spain.}

\end{center}

\vskip 3ex

\begin{abstract}

\vspace{0.5cm}
%\begin{center}

In this paper we discuss a connection between Dirac operators on configuration spaces and Yang-Mills quantum field theory. We first show that the Hamilton operators of the self-dual and anti-self-dual sectors of a Yang-Mills quantum field theory emerge from unitary transformations of a Dirac equation formulated on a configuration space of gauge connections.  Secondly, we formulate a Bott-Dirac operator on the configuration space and demonstrate how the Hamilton operator of a Yang-Mills quantum field theory coupled to a fermionic sector emerges from its square. Finally, we discuss a spectral invariant that emerges in this framework. %If the Dirac or Bott-Dirac operators are well-defined then so are these Yang-Mills Hamilton operators.

%\end{center}
\end{abstract}

\newpage
\section{Introduction}

One of the most important unsolved problems in contemporary theoretical high-energy physics is the rigorous formulation of non-perturbative quantum Yang-Mills theory \cite{Jaffe:2000ne}. %This problem has remained unsolved for more than half a century. 
The fact that this problem has remained unsolved for more than half a century suggests that the correct framework for addressing it has not yet been found and, as a consequence, that it has not yet been fully understood what quantum Yang-Mills theory actually is.

In this short paper we propose a new approach to this problem by demonstrating a connection between Dirac operators on configuration spaces and non-perturbative quantum Yang-Mills theory. Firstly, we assume that a Dirac equation on a configuration space can be rigorously defined and then we show that unitary transformations of this equation leads to the Hamiltonians of the self-dual and anti-self-dual sectors a quantum Yang-Mills theory. The Hamilton operator emerges from the square of the Dirac operator and the unitary transformation involves the Chern-Simons term. Interestingly, the unitary transformation produces an additional term, which is a spectral invariant of a covariant derivative on the underlying manifold.

Secondly, we formulate a Bott-Dirac operator on the configuration space and find that in addition to the Hamiltonian of a Yang-Mills quantum field theory its square also generates the Hamilton operator of a fermionic sector that involves a quantised fermionic field. This Bott-Dirac operator is a gauge-covariant variant of the Bott-Dirac operator constructed by Higson and Kasparov in \cite{Higson}.

This paper is the result of an ongoing research project \cite{Aastrup:2023jfw}-\cite{Aastrup:2019yui}
 concerned with a rigorous formulation of a geometrical construction on a configuration space of gauge connections using noncommutative geometry \cite{ConnesBook,1414300}. In previous publications we have shown that the key building blocks of a Yang-Mills quantum field theory coupled to a fermionic sector on a curved background \cite{Aastrup:2020jcf}, i.e. the Hamilton operators and the canonical commutation and anti-commutation relations, emerge from a spectral triple-like construction on a configuration space. In those papers we did not consider unitary transformations of the Dirac operator on the configuration space.

The notion of a geometry of configuration spaces of gauge connections was considered already by Feynman \cite{Feynman:1981ss} and Singer \cite{Singer:1981xw} (see also \cite{Orland:1996hm}). The idea to study non-trivial geometries and in particular to study their dynamics, is, however, new.

In order to enhance readability and to emphasise the simplicity of the results presented we have kept the level of detail at a minimum in this paper. We refer the reader to \cite{Aastrup-newpaper,Aastrup:2023jfw} for a more rigorous treatment.

%[EVERYWHERE: $\frac{\pa}{\pa x_i}$ instead of $\frac{\pa}{\pa \xi_i}$]

\section{A Dirac equation on a configuration space}

Let $\ca$ be a configuration space of $SU(2)$ gauge connections on a three-dimensional manifold $M$. 
We denote by $\{\xi_i\}$ an orthonormal basis of $\ca$ and write $\ca\ni A(m)=\sum_i x_i \xi_i(m)$. With this notation we define a derivative on $\ca$ as
$$
\frac{\d}{\d A (m)} = \sum_i \xi_i(m) \frac{\pa}{\pa x_i} 
$$
that satisfies
\begin{equation}
\left[ \frac{\d}{\d A(m_1)} , A(m_2)\right] = \sum_i \xi_i(m_1)\xi_i(m_2) = \d^3(m_1-m_2)
\label{K}
\end{equation}
where the last equality only holds when the basis vectors $\xi_i$ are orthogonal with respect to an $L^2$-norm. We shall assume that this is the case throughout this paper\footnote{Note that this need not be the case: it is possible that the setup involves a non-locality, either through an UV-regularisation or through some other effect, in which case the integral kernel $\sum_i \xi_i(m_1)\xi_i(m_2)$ will be more involved than the Dirac delta function. See \cite{Aastrup:2023jfw,Aastrup:2020jcf} for details.}.

Next, we introduce Clifford elements
$\bar{c}_i $ that satisfy the anti-commutator relation 
\begin{equation}
\{\bar{c}_i,\bar{c}_j\} = - \d_{ij}
\label{rel1}
\end{equation}
and use them to define the Dirac operator 
\begin{equation}
D = \sum_i \bar{c}_i \frac{\pa}{\pa x_i}.
\label{RFK}
\end{equation}
We shall assume that a Hilbert space $$\ch=L^2(\ca)\otimes \ch_f,$$
where $\ch_f$ is a fermionic Fock space, exists and that $D$ is a well-defined operator in $\ch$.
With this setup we are finally ready to write down a Dirac equation on $\ca$
\begin{equation}
D \Psi(A) = 0.
\label{Deq}
\end{equation}

\section{A unitary transformation}

Next we consider the unitary operator 
$$
U = \exp (i k CS(A))
$$
where $k$ is an integer divided by $4\p$ and where $CS(A)$ is the Chern-Simons term
$$
CS(A) = \int_M \mbox{Tr} \left( {A}\wedge d{A} + \frac{2}{3} {A}\wedge {A} \wedge {A}\  \right),
$$
where the trace is over the Lie-algebra of $SU(2)$, and write down the rotated version of the Dirac equation (\ref{Deq})
$$
D^{\mbox{\tiny $U$}} \Psi^{\mbox{\tiny $U$}}(A) = 0
$$
where $\Psi^{\mbox{\tiny $U$}}(A) = U \Psi(A)$ and where
\begin{eqnarray}
D^{\mbox{\tiny $U$}} = U D U^* =                %\nn\\
D - [D,U]U^*.
\label{fluk}
\end{eqnarray}
We can write the second term in (\ref{fluk}) as 
$$
[D,U]U^* 
%= i k \sum_i \bar{c}_i \frac{\pa CS}{\pa x_i} 
= 2i k \sum_i \bar{c}_i \int_M \mbox{Tr}\left(\xi_i \wedge F(A)  \right)
$$
where $F(A)$ is the field strength tensor of the connection $A$, and where we used that
$$
\frac{\pa CS}{\pa x_i} = 2 \int_M \mbox{Tr}\left(\xi_i \wedge F(A)  \right).
$$
We then compute the square of $D^{\mbox{\tiny $U$}}$ 
%[remove all but the last term]
\begin{eqnarray}
    \left( D^{\mbox{\tiny $U$}}\right)^2 
%    
%    
%    &=&
%UDU^* UDU^* = 
%U D^2 U^*
%\nn\\&=&
%-e^{ik CS(A)} \sum_{i}\left(\frac{\pa}%{\pa x_i} \right)^2  e^{-ik CS(A)}
%\nn\\ &=&
%-e^{ik CS(A)} \sum_{i}\frac{\pa}{\pa x_i} % \left(-ik \frac{\pa CS(A)}{\pa x_i}e^{-%ik CS(A)} + e^{-ik CS(A)} \frac{\pa}{\pa %x_i}\right)  
%\nn\\
%&=&
%-e^{ik CS(A)} \sum_{i} \left(-ik %\frac{\pa^2 CS(A)}{\pa x_i \pa x_i}e^{-ik %CS(A)} 
%- k^2 \frac{\pa CS(A)}{\pa x_i}\frac{\pa %CS(A)}{\pa x_i} e^{-ik CS(A)}\right.
%\nn\\&&\left.
%-ik \frac{\pa CS(A)}{\pa x_i}e^{-ik %CS(A)}\frac{\pa}{\pa x_i}
%-ik \frac{\pa CS(A)}{\pa x_i} e^{-%ikCS(A)}\frac{\pa}{\pa x_i}
%+ e^{-ik CS(A)} \left(\frac{\pa}{\pa %x_i}\right)^2 \right)  
%\nn\\
%
%
&=& k^2\sum_{i} \Bigg(\frac{i}{k} \frac{\pa^2 CS(A)}{\pa x_i \pa x_i} 
+ \left( \frac{\pa CS(A)}{\pa x_i}\right)^2 
\nn\\&&
\qquad \quad+\frac{2i}{k} \frac{\pa CS(A)}{\pa x_i} \frac{\pa}{\pa x_i}
-  \frac{1}{k^2}\left(\frac{\pa}{\pa x_i}\right)^2 \Bigg).  
\label{hovedpine22}
\end{eqnarray}
In order to interpret equation (\ref{hovedpine22}) we introduce the notation 
$$
\hat{E}_i := \frac{i}{2k}\frac{\pa}{\pa x_i},\quad \hat{E}(m) = \sum_i \xi_i(m)\hat{E}_i 
$$
and use 
\begin{equation}
 \frac{\pa^2 CS(A)}{\pa x_{i } \pa x_{j }} 
 =    \int_M \mbox{Tr} \left(  \xi_{i  }  \wedge \nabla^A  \xi_{j} \right) + \int_M \mbox{Tr} \left(  \xi_{j  }  \wedge \nabla^A  \xi_{i} \right) ,
 \label{second}
\end{equation}
to write it as
\begin{eqnarray}
    \left( D^{\mbox{\tiny $U$}}\right)^2 
=
H^{(-)} 
\nn\label{hovedpine2}
\end{eqnarray}
with
\begin{eqnarray}
    H^{(\pm)} = 4k^2\left(  \left( \hat{E}_i \right)^2 +  \int_M \mbox{Tr}\left( F(A)^2\right) 
\pm   2  \int_M \mbox{Tr}\left( F(A) \wedge\hat{E}  \right)
%\right.\nn\\&&\left.
\mp \mbox{Tr}_\xi \left(i\nabla^A \right)\right).
\nn\label{simonn}
\end{eqnarray}
Here we used that
\begin{equation}
\left( \frac{\pa CS(A)}{\pa x_i}\right)^2 = \int_M \mbox{Tr}\left( F(A)^2 \right),
\nn
\end{equation}
which holds whenever the integral kernel in (\ref{K}) gives us the Dirac delta-function. Also, the last term in (\ref{simonn}) is the trace of $i\nabla^A$, which is a spectral invariant
$$
\mbox{Tr}_\xi \left(i\nabla^A \right)=\sum_i\frac{i}{4k}\int_M \mbox{Tr} \left(  \xi_{i  }  \wedge \nabla^A  \xi_{i} \right)
$$
 that we first discussed a variant of in section 5.3 in \cite{Aastrup:2020jcf} and which we shall discuss shortly.

Note that if we instead of $U$ use $U^*$ in (\ref{fluk}) we obtain
\begin{eqnarray}
    \left( D^{\mbox{\tiny $U^*$}}\right)^2 
=
H^{(+)} .
\nn\label{hovedpine3}
\end{eqnarray}

Let us compare $H^{(\pm)}$ to a Langrangian setup. If we write the Yang-Mills action as
\begin{eqnarray}
S_{\mbox{\tiny YM}}
= S^{(+)}_{\mbox{\tiny YM}} + S^{(-)}_{\mbox{\tiny YM}} 
\nn%\label{Spm}
\end{eqnarray}
with
$$
S^{(\pm)}_{\mbox{\tiny YM}}=\frac{1}{2}\int_\cm \mbox{Tr}\left( {\bf F}\wedge \left( \star {\bf F} \pm \theta {\bf F}\right)\right)
$$
where $\cm$ is now a four-dimensional manifold in which $M$ is a Cauchy surface, ${\bf F}$ is a four-dimensional field-strength tensor, and $\star$ is the four-dimensional Hodge dual, then we see that $H^{(\pm)}$ are the Hamiltonians that correspond to the selfdual and anti-selfdual sectors of an $SU(2)$ Yang-Mills theory in the sense that
$$
H_{\mbox{\tiny YM}} = H^{(+)}+H^{(-)}.
$$

\section{A fermionic sector}

Let us now consider instead of the Dirac operator (\ref{RFK}) a Bott-Dirac operator. To do this we need the full Clifford algebra,
$$
c_i = a_i^\dagger + a_i,\quad \bar{c}_i = a_i^\dagger - a_i
$$
where $(a^\dagger_i, a_i)$ are respectively the creation and annihilation operators acting in $\ch_f$. In addition to (\ref{rel1}) we have the relations
$$
\{c_i,c_j\}=\d_{ij},\quad \{\bar{c}_i,c_j\}=0.  
$$
With this we define the Bott-Dirac operator as
\begin{equation}
B = \sum_i \left( \bar{c}_i \frac{\pa}{\pa x_i} + \k c_i \frac{\pa CS(A)}{\pa x_i} \right),
\label{Bott}
\end{equation}
where $\kappa$ is a real constant, and compute its square
$$
B^2 = \sum_i\left( - \left( \frac{\pa}{\pa x_i}\right)^2
+ \k^2 \left( \frac{\pa CS(A)}{\pa x_i}\right)^2\right)
+ \k \sum_{ij} \bar{c}_i c_j \frac{\pa CS(A)}{\pa x_i \pa x_j} .
$$
We recognize the first two terms as the Hamiltonian for an $SU(2)$ Yang-Mills theory. Using (\ref{second}) we can rewrite the second term as
$$
 \sum_{ij} \bar{c}_i c_j \frac{\pa CS(A)}{\pa x_i \pa x_j}
=     \int_M \mbox{Tr} \left(  
%{\psi}^\dagger  \wedge \nabla^A  \psi^\dagger 
%-   \psi  \wedge \nabla^A  {\psi} 
%+
\hat{\phi}^\dagger  \wedge \nabla^A  \hat{\phi} 
-   \hat{\phi}  \wedge \nabla^A  \hat{\phi}^\dagger \right)
$$
where we introduced the fermionic field operators
$$
\hat{\phi}(m) = \sum_i a_i \xi_i(m),\quad \hat{\phi}^\dagger(m) = \sum_i a^\dagger_i \xi_i(m),
$$
which satisfy the relation
\begin{equation}
\{\hat{\phi}^\dagger (m_1),\hat{\phi}(m_2)\} = \sum_i \xi_i(m_1)\xi_i(m_2).
\label{third}
\end{equation}
If we once more assume that our construction is local (or take its local limit) then (\ref{third}) is the canonical anti-commutation relation of a quantised fermionic field.
Note, however, that the fermionic fields $(\hat{\phi},\hat{\phi}^\dagger)$ are one-forms. Also, since we have not specified how the Clifford elements $(c_i,\bar{c}_j)$ are constructed we cannot say what spin they have. We refer the reader to \cite{Aastrup-newpaper} where we have constructed a framework that involves half-integer fermions.

To summarize, we find that the square of the Bott-Dirac operator (\ref{Bott}) gives us two terms
\begin{eqnarray}
    B^2 = H_{\mbox{\tiny YM}} + H_{\mbox{\tiny fermionic}}
    \nn
\end{eqnarray}
where 
\begin{eqnarray}
H_{\mbox{\tiny YM}} &=& 4{\kappa^2}  \left( \left( k \hat{E}_i \right)^2 +  \int_M \mbox{Tr}\left( F(A)^2\right) \right)
\nn\\
H_{\mbox{\tiny fermionic}} &=& \kappa \left(    \int_M \mbox{Tr} \left(  
\hat{\phi}^\dagger  \nabla^A  \hat{\phi} 
-   \hat{\phi}   \nabla^A  \hat{\phi}^\dagger \right)
\right)
\nn
\end{eqnarray}
is respectively the Hamiltonian of an $SU(2)$ Yang-Mills quantum gauge theory and
the principal part of a Dirac Hamiltonian of a fermionic sector.

\section{The spectral invariant}

Before we finish let us take a brief look at the Hamiltonian $H_{\mbox{\tiny fermionic}}$. If we bring it on a normal-ordered form, where the annihilation operators stand to the right
$$
H_{\mbox{\tiny fermionic}} = 2\kappa     \int_M \mbox{Tr} \left(  
\hat{\phi}^\dagger  \nabla^A  \hat{\phi} 
\right) + \mbox{Tr}_{\phi} \left( \nabla^A\right)
$$
we find once more the operator-trace
\begin{equation}
\mbox{Tr}_{\phi} \left(\nabla^A\right) = -\kappa \sum_i \int_M \mbox{Tr} \left( \phi_i \nabla^A \phi_i \right).
\label{hugo}
\end{equation}
Equation (\ref{hugo}) is the spectral invariant that we first discussed in section 5.3 in \cite{Aastrup:2020jcf} and which is related to the eta-invariant for $\nabla^A$ that was first introduced by Atiyah, Patodi, and Singer \cite{Atiyah}-\cite{AtiyahIII}. The spectral invariant, which measures the assymmetry of the spectrum of $\nabla^A$, plays here the role of a vacuum energy for the fermionic sector. In \cite{Aastrup:2020jcf} we showed that $\mbox{Tr}_{\phi} \left(\nabla^A\right)$ exist independently of any UV-regularisation one might use to compute it.\\

%The term $\Tr_\psi (D^A)$ corresponds to the %contribution
%$$
%\int d^3 p E_p \d^{(3)}(0)  
%$$
% that emerges in ordinary fermionic quantum field theory and which is usually removed using normal ordering. 

Finally, the ground state of the Bott-Dirac operator $B$ is
$$
\Psi'(A) = e^{-\kappa CS(A)}\otimes \vert 0 \rangle, \quad B\Psi' =0,
$$
where $\vert 0 \rangle$ is the zero-particle state in $\ch_f$. Thus, in this case we do not find a complex phase but instead a real Chern-Simons term. From a mathematical point of view this is more challenging to handle since it puts stronger demands on the construction of the Hilbert space $L^2(\ca)$.
Note also that whereas the ground state in (\ref{Deq}) has a degeneracy that depends on precisely how the Dirac operator on $\ca$ is constructed\footnote{Technically, the construction of the Dirac operator $D$ requires a metric on $\ca$ and hence a corresponding covariant derivative in $D$. If we have the trivial geometry on $\ca$ the ground state will be infinitely degenerate, but for non-trivial geometries this degeneracy may be somewhat removed. }, the ground state for $B$ has no degeneracy.

\section{Discussion}

The results presented in this paper are interesting for several reasons. 
First of all, since Dirac operators are intrinsically geometrical objects, our result suggest that in order to understand quantum Yang-Mills theory one needs to consider the geometry of their configuration spaces, and in particular, to consider their {\it dynamics}. In other words, this suggests that one should apply the machinery of Einsteins theory of relativity to the central playing field of quantum field theory, which is the configuration space, and consider nontrival, dynamical geometries. This is a paradigmatically different from other approaches to fundamental physics seen today. In \cite{Aastrup:2005yk} we have initiated such a programme.

Secondly, the presence of Dirac operators points in the direction of noncommutative geometry \cite{ConnesBook,1414300,Connes:1996gi}, as does the fact that the natural algebra to consider in this setup, the $\mathbf{HD}$-algebra \cite{Aastrup:2012vq,AGnew}, which is generated by parallel transports along flows of vectorfields on the underlying manifold, is inherently noncommutative.
The reason why this is interesting is that noncommutative geometry provides a rich mathematical machinery that involves several mechanisms of unification. We already demonstrated how a Bott-Dirac operator gives us a fermionic sector, but note also that the term
\begin{equation}
[D,U]U^*,
\nn\label{oone}
\end{equation}
which was added to the Dirac operator when we rotated it unitarily, in the terminology of noncommutative geometry is the simplest example of a one-form on the configuration space \cite{ConnesBook,Connes:1996gi}. That is, within this framework it is natural to consider gauge theories {\it on} the space of gauge connections (a truly meta-concept), which will then lead to more interesting structures that could be important in a semi-classical/low-energy limit.
Thus, the conclusion that emerges from all this is that nonperturbative Yang-Mills theory could be much more than merely Yang-Mills theory quantised: it could lead to a theory of unification.

%Furthermore, the one-form in (\ref{oone}) contains a term that has a double derivative of the Chern-Simons term, which gives the covariant derivative on the underlying manifold. In the current setup this terms gives us the operator-trace of the covariant derivative, but if one were instead to construct a Bott-Dirac operator on the configuration space then this term gives a Hamiltonian \cite{} of a fermionic sector that comes with the fermionic Fock space that is needed in order to have a representation of the infinite-dimensional Clifford algebra used to construct the Dirac operator. 

%Finally, it is an interesting question what role the spectral invariant $\mbox{Tr}_\xi \left(i\nabla^A \right)$ that turns up as an additional term in the Yang-Mills Hamiltonian might play. One obvious interpretation is that if it is positive it will provide a mass gap in the emerging Yang-Mills theory. Another, and not mutually exclusive interpretation, is that it could be related to the cosmological constant. Note, however, that since it depends on the connection $A$ it will have a time-evolution.

There are many technical challenges involved in rigorously defining a Dirac operator on a configuration space as we have previously discussed \cite{Aastrup:2023jfw}-\cite{Aastrup:2019yui}. For instance, issues with gauge fixing and in particular the Gribov ambiguity \cite{Gribov:1977wm} must be addressed, as well as the question of existence and convergence. We have, however, shown that it is possible to rigorously define a Dirac operator on a configuration space of gauge connections in certain cases \cite{Aastrup:2023jfw}.

\vspace{0,7cm}
\noindent{\bf\large Acknowledgements}\\

\noindent

JMG would like to express his gratitude to entrepreneur Kasper Bloch Gevaldig for his unwavering financial support. JMG is also indepted to Master of Science in Engineering Vladimir Zakharov and to the following list of sponsors for their generous support:   Frank Jumppanen Andersen,
Bart De Boeck, Simon Chislett,
Jos van Egmond,
Trevor Elkington,
Jos Gubbels,
Claus Hansen,
David Hershberger,
Ilyas Khan,
Simon Kitson,
Hans-J\o rgen Mogensen,
Stephan M{\"u}hlstrasser,
Bert Petersen,
Ben Tesch,
Jeppe Trautner,
and the company
Providential Stuff LLC. JMG would also like to express his gratitude to the Institute of Analysis at the Gottfried Wilhelm Leibniz University in Hannover, Germany, for kind hospitality during numerous visits.

\end{document}